\documentclass[aps,groupedaddress,nofootinbib,floatfix,epfs,preprintnumbers]{revtex4}
\usepackage{amsfonts,amscd,amsmath,amssymb,graphicx,color,float}
\usepackage[section]{placeins}
\def\ep{\text{e}}
\def\g{\mathsf{g}}
\def\oh{\frac{1}{2}}
\def\s{\mathsf{s}}

\def\n{\mathsf{n}}

\def\rh{r_h}

\def\rq{r_q}

\def\rqb{r_{\bar q}}

\def\Vc{V_\text{\scriptsize con}}
\def\Vd{V_\text{\scriptsize dis}}
\begin{document}
%\preprint{LMU-ASC ??/26}
\title{The spatial Wilson loops, string breaking, and AdS/QCD}
\author{Oleg Andreev}
 %\affiliation{L.D. Landau Institute for Theoretical Physics, Kosygina 2, 119334 Moscow, Russia}
  %\affiliation{V.A. Steklov Mathematical Institute, Gubkina 8, 119991, Moscow, Russia}
\thanks{Also on leave from L.D. Landau Institute for Theoretical Physics}
\affiliation{Arnold Sommerfeld Center for Theoretical Physics, LMU-M\"unchen, Theresienstrasse 37, 80333 M\"unchen, Germany}
%\date{}
\begin{abstract} 
 We consider the phenomenon of string breaking in the context of the spatial Wilson loops using the gauge/string duality. In particular, we discuss the impact of light flavors on the pseudopotential. We also introduce the notion of the spatial string breaking distance and estimate it for $SU(3)$ gauge theory in the temperature range $0\,\text{-}\,3\,T_c$.
 \end{abstract}
%\pacs{empty}
\maketitle
%__________________________________________________________ section 2
\section{Introduction}
\renewcommand{\theequation}{1.\arabic{equation}}
\setcounter{equation}{0}

It is well-known that QCD at high temperature undergoes a phase transition from a confined phase to a deconfined phase.\footnote{For $2+1$ flavor QCD with physical quark masses, this transition is likely a rapid crossover. See, for example, \cite{aoki}.} While basic thermodynamical observables show a drastic qualitative change at the transition point, there are some whose structure does not change qualitatively across $T_c$. One example is the pseudopotential $V$ extracted from spatial Wilson loops. For a rectangular loop ${\cal C}$ of length $\ell$ and width $Y$ along two spatial directions, it is defined as 

\begin{equation}\label{Wlattice}
\langle W({\cal C})\rangle \sim \ep^{-V(\ell) Y}
\,,\qquad\text{as}\qquad Y\to \infty
\,.
\end{equation}
Importantly, at any temperature the spatial Wilson loops obey an area law (for large $\ell$ and $Y$), so that $V=\sigma_s\ell$, where $\sigma_s$ is the spatial string tension \cite{borgs}. This implies the confinement of magnetic modes.

For temperatures below and just above $T_c$, the computation of the pseudopotential is strongly influenced by nonperturbative effects and therefore cannot be performed within perturbative QCD. Although lattice gauge theory is one of the basic tools for studying nonperturbative phenomena and has made significant progress in calculating the spatial Wilson loops in four dimensions \cite{Vlattice}, the need to understand the physics behind computational complexity motivates the use of effective field and string models. A special class of string models, known as AdS/QCD (holographic) models, has attracted much attention in the recent years. The hope is that the gauge/string duality provides new theoretical tools for studying strongly coupled gauge theories.\footnote{For the further development of these ideas in the context of QCD, see the book \cite{book-u} and references therein.} 

This Letter extends our study of the pseudopotentials in pure gauge theories \cite{az2} to theories with two light quarks. The phenomenon of string breaking plays a pivotal role in this, as it modifies the large-$\ell$ behavior of the pseudopotential. 

%___________________________________________________________
\section{Key features of a five-dimensional framework}
\renewcommand{\theequation}{2.\arabic{equation}}
\setcounter{equation}{0}

In the context of AdS/CFT (QCD), the discussion of a Wilson loop proceeds as follows. First, one chooses a contour ${\cal C}$ on a four-manifold which is the boundary of a five-dimensional manifold. One then considers fundamental strings on this manifold such that the string worldsheet has ${\cal C}$ as its boundary. The expectation value of the Wilson loop is schematically given by the worldsheet path integral 

\begin{equation}\label{wilson}
\langle\,W({\cal C})\,\rangle=\int DX\,\ep^{-S_{\text w}}
\,,
\end{equation}
where $X$ denotes a set of worldsheet fields and $S_{\text w}$ is a worldsheet action. In principle, this integral can be evaluated semiclassically in terms of minimal surfaces satisfying the boundary conditions. The result takes the form

\begin{equation}\label{wilson2}
\langle\,W({\cal C})\,\rangle=\sum_n w_n\ep^{-S_n}
\,,
\end{equation}
where $S_n$ represents a regularized minimal area whose relative weight is $w_n$.\footnote{The key point is that these areas are divergent, but the 
divergences are proportional to the circumference of ${\cal C}$.}

To illustrate these ideas, we now consider a specific model. This model is well motivated for two reasons: first, it provides the estimates for Wilson loops which  agree well with lattice calculations and QCD phenomenology in four dimensions \cite{az2,white,az4}; second, its simplicity enables many analytical estimates.

Following \cite{az2}, we consider a five-dimensional metric which is a simple one-parameter deformation of the Schwarzschild black hole in Euclidean $\text{AdS}_5$. Light quarks are modeled by turning on a scalar field $\text T$ \cite{son}, which accounts for quarks at the string endpoints. Thus, the background takes the form

\begin{equation}\label{metric}
ds^2=\ep^{\s r^2}\frac{R^2}{r^2}\Bigl(f(r)dt^2+d\vec x^2+f^{-1}(r)dr^2\Bigr)
\,,
\qquad
{\text T}={\text T}(r)
	\,.
\end{equation}
Here $f(r)=1-\frac{r^4}{\rh^4}$ and $\s$ is the deformation parameter. The boundary is located at $r=0$. The Hawking temperature, which is identified with the temperature of the dual gauge theory, is $T=\frac{1}{\pi\rh}$. The critical temperature is determined from the spatial string tension \cite{az2}. Explicitly,

\begin{equation}\label{Tmu-pc}
T_c=\frac{\sqrt{\s}}{\pi}
\,.
\end{equation}

To construct string configurations in this background, we need two key ingredients. The first is a Nambu-Goto string governed by the action 

\begin{equation}\label{NG}
S_{\text{\tiny NG}}=\frac{1}{2\pi\alpha'}\int d^2\xi\,\sqrt{\gamma^{(2)}}
\,,
\end{equation}
where $\gamma$ is an induced metric, $\alpha'$ is a string parameter, and $\xi^i$ are worldsheet coordinates. The second is a boundary term 

\begin{equation}\label{Sq}
S_{\text{q}}=\int d\tau e\,\text{T}
\,,
\end{equation}
which is the usual sigma-model action for strings in a tachyon background. The integral is over a worldsheet boundary, parameterized by $\tau$, and $e$ is a boundary metric. 
%________________________________________________________________
\section{The pseudopotential}
\renewcommand{\theequation}{3.\arabic{equation}}
\setcounter{equation}{0}

\begin{figure}[htbp]
\centering
\includegraphics[width=5.5cm]{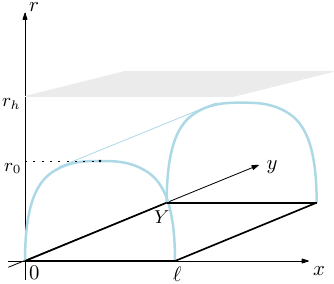}
\hspace{2.75cm}
\includegraphics[width=5.5cm]{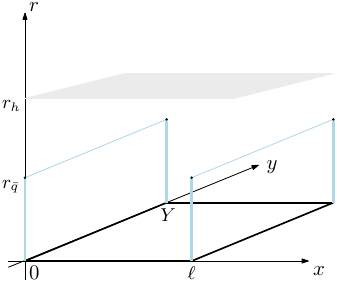}
\caption{{\small String configurations in five dimensions. The rectangle at $r=0$ is shown in bold and the horizon at $r=\rh$ in gray. The string profiles are sketched at $y=0$ and $y=Y$. Left: A connected configuration. Here $r_0$ is the $r$-coordinate of the string turning point. Right: A disconnected configuration. Here $\rqb$ denotes a $r$-coordinate of the light quarks.}}
\label{confs}
\end{figure}

Now we want to compute the pseudopotential in the presence of light dynamical quarks. Since $V$ coincides with the heavy quark-antiquark potential at zero temperature, it is natural to consider not only the connected string configuration but also the disconnected one \cite{astb}. The latter can be interpreted as a pair of heavy-light mesons created by the formation of a virtual quark-antiquark pair. Both configurations are shown in Fig.\ref{confs}.\footnote{For computing the ground state pseudopotential, there is no need to involve excited strings, which lead to hybrid pseudopotentials. See, for example, \cite{ahybrid}.}
%________________________  fig - 1  ____________________________

%___________________________________________________________________________________
\subsection{The connected configuration}

The pseudopotential related to the connected configuration can be computed along the lines described in \cite{malda}. In the limit $Y\to\infty$, the area swept out by the string is obtained by parallel transporting it along the $y$-direction (see Fig.\ref{confs}), yielding the pseudopotential $\Vc=S_{\text{\tiny NG}}/Y$. An important point is that the side areas at $y=0$ and $y=Y$ become irrelevant in this limit. For the geometry \eqref{metric}, the result is expressed parametrically \cite{az2}  

\begin{equation}\label{El}
\begin{split}	
\ell=&2\sqrt{\frac{\lambda}{\s}}	
\int_0^1 dv \,v^2\ep^{\lambda(1-v^2)}
\biggl(1-\lambda^2v^4\frac{T^4}{T_c^4}\biggr)^{-\oh}
\Bigl(1-v^4\ep^{2\lambda(1-v^2)}\Bigr)^{-\oh}
\,,\\
\Vc=&2\g\sqrt{\frac{\s}{\lambda}}
\int_0^1\frac{dv}{v^2}
\biggl[
\ep^{\lambda v^2}
\biggl(1-\lambda^2v^4\frac{T^4}{T_c^4}\biggr)^{-\oh}
\Bigl(1-v^4\ep^{2\lambda(1-v^2)}\Bigr)^{-\oh}
-
1-v^2
\biggr]
+2c
\,.
\end{split}
\end{equation}
Here $c$ is a normalization constant arising from the renormalization of a linear divergence. The parameter $\lambda$ is defined as $\lambda=\s r_0^2$. It varies from $0$ to $1$ for $T\leq T_c$ and from $0$ to $\frac{T_c^2}{T^2}$ for $T\geq T_c$. This reflects the underlying two-wall pattern, as explained in \cite{az2}.\footnote{A similar pattern persists in the anisotropic case \cite{IA}.} The first wall is a soft wall at $r=1/\sqrt{\s}$, which confines the string to the near-boundary region for $T<T_c$, while the second is the horizon, which confines it for $T>T_c$. The walls coincide at $T=T_c$. This pattern qualitatively explains the temperature dependence of the spatial string tension which is temperature-independent below $T_c$ and manifestly dependent above $T_c$.\footnote{We call such behavior "specific" even if it exhibits weak temperature dependence below $T_c$.} In the current model, the spatial string tension is given by 

\begin{equation}\label{large-l}
\sigma_s=
\begin{cases}
\sigma & \text{if} \quad T\leq T_c\,,\\
\sigma\frac{T^2}{T_c^2}\exp\bigl\lbrace \frac{T_c^2}{T^2}-1\bigr\rbrace 
& \text{if} \quad T\geq T_c\,,
\end{cases}
	\end{equation}
where $\sigma=\g\ep\s$ is the physical string tension at zero temperature \cite{az1}.     

For our purposes here, we need to know the subleading term in the large-$\ell$ expansion of the pseudopotential. To find it, consider the difference

\begin{equation}\label{large-l2}
	\begin{split}
	\Vc-\sigma_s\ell
	&=2\g\sqrt{\s}
	\biggl(\frac{1}{\sqrt{\lambda}}
\int_0^1\frac{dv}{v^2}
\biggl[
\ep^{\lambda v^2}
\biggl(1-\lambda^2v^4\frac{T^4}{T_c^4}\biggr)^{-\oh}
\Bigl(1-v^4\ep^{2\lambda(1-v^2)}\Bigr)^{-\oh}
-
1-v^2
\biggr]\\
&\quad-\sqrt{\lambda}
\int_0^1 dv \,v^2\ep^{1+\lambda(1-v^2)}
\biggl(1-\lambda^2v^4\frac{T^4}{T_c^4}\biggr)^{-\oh}
\Bigl(1-v^4\ep^{2\lambda(1-v^2)}\Bigr)^{-\oh}
\biggr)
+
2c
\,,
\end{split}
\end{equation}
where $T<T_c$. Taking the limit $\lambda\to 1$ yields
 
 \begin{equation}\label{large-l3}
 	\Vc-\sigma_s\ell=
 	2\g\sqrt{\s}
\int_0^1\frac{dv}{v^2}
\biggl[
\ep^{v^2}
\biggl(1-v^4\frac{T^4}{T_c^4}\biggr)^{-\oh}
\Bigl(1-v^4\ep^{2(1-v^2)}\Bigr)^{\oh}
-
1-v^2
\biggr]
+2c
\,.
 \end{equation}
Similarly, for $T>T_c$, the difference can be written as

\begin{equation}\label{large-l4}
	\begin{split}
	\Vc-\sigma_s\ell
	&=2\g\sqrt{\s}
	\biggl(\frac{1}{\sqrt{\lambda}}
\int_0^1\frac{dv}{v^2}
\biggl[
\ep^{\lambda v^2}
\biggl(1-\lambda^2v^4\frac{T^4}{T_c^4}\biggr)^{-\oh}
\Bigl(1-v^4\ep^{2\lambda(1-v^2)}\Bigr)^{-\oh}
-
1-v^2
\biggr]\\
&\quad-\sqrt{\lambda}\frac{T^2}{T_c^2}\ep^{\tfrac{T_c^2}{T^2}}
\int_0^1 dv \,v^2\ep^{\lambda(1-v^2)}
\biggl(1-\lambda^2v^4\frac{T^4}{T_c^4}\biggr)^{-\oh}
\Bigl(1-v^4\ep^{2\lambda(1-v^2)}\Bigr)^{-\oh}
\biggr)
+
2c
\,.
\end{split}
\end{equation}
Taking the limit $\lambda\to\tfrac{T_c^2}{T^2}$  then gives

\begin{equation}\label{large-l5}
	\Vc-\sigma_s\ell
	=
	2\g\sqrt{\s}\frac{T}{T_c}\int_0^1\frac{dv}{v^2}
\biggl[
\ep^{\frac{T_c^2}{T^2} v^2}
\bigl(1-v^4\bigr)^{-\oh}
\Bigl(1-v^4\ep^{2\frac{T_c^2}{T^2}(1-v^2)}\Bigr)^{\oh}
-
1-v^2
\biggr]
+2c
\,.
\end{equation}
Putting the two pieces together, the large-$\ell$ expansion of $V$ reads 

\begin{equation}\label{large-l6}
\Vc=\sigma_s\ell+C+o(1)
\,,\quad\text{with}\quad
C=
\begin{cases}
2c-2\g\sqrt{\s}\int_0^1\frac{dv}{v^2}\Bigl[1+v^2-
\ep^{v^2}\Bigl(1-v^4\frac{T^4}{T_c^4}\Bigr)^{-\oh}\bigl(1-v^4\ep^{2(1-v^2)}\bigr)^\oh\Bigr]
& \text{if} \quad T\leq T_c\,,\\
2c-2\g\sqrt{\s}\frac{T}{T_c}
\int^1_0
\frac{dv}{v^2}\Bigl[1+v^2-
\ep^{v^2\frac{T_c^2}{T^2}}\bigl(1-v^4\bigr)^{-\oh}\bigl(1-v^4\ep^{2\frac{T_c^2}{T^2}(1-v^2)}\bigr)^\oh\Bigr]
& \text{if} \quad T\geq T_c\,.
\end{cases}
	\end{equation}

To complete the picture, we now consider the small-$\ell$ behavior of the pseudopotential, which corresponds to the limit $\lambda\to 0$. Keeping only the leading terms in \eqref{El}, we obtain 

\begin{equation}\label{small}
\ell=\oh\sqrt{\frac{\lambda}{\s}}
\bigl(B\bigl(\tfrac{3}{4},\tfrac{1}{2}\bigr)+O(\lambda)\bigr)
	\,,\qquad
	\Vc=\oh\g\sqrt{\frac{\s}{\lambda}}\bigl(B\bigl(-\tfrac{1}{4},\tfrac{1}{2}\bigr)+O(\lambda)\bigr)+2c
\,,
\end{equation}
where $B(a,b)$ is  the beta function. From this, it immediately follows that 

\begin{equation}\label{small2}
\Vc=-\frac{\alpha}{\ell}+2c+o(1)
	\,,\qquad\text{with}\qquad
	\alpha=\g\,\frac{(2\pi)^3}{\Gamma^4\bigl(\tfrac{1}{4}\bigr)}
	\,.
\end{equation}
Notably, to this order, $\Vc$ is independent of temperature and coincides with the heavy quark-antiquark potential at zero temperature \cite{malda,az1}.  

%_________________________________________________________________________
\subsection{The disconnected configuration}

The disconnected string configuration is shown on the right panel of Fig.\ref{confs}. The motivation for considering such a configuration comes from the construction of the heavy quark-antiquark potential in the presence of light quarks \cite{astb}, where the disconnected configuration describes a pair of heavy-light mesons formed in the decay process $Q\bar Q\to Q\bar q +q\bar Q$. At zero quark chemical potential it consists of two identical parts. The areas swept out by the strings are obtained by parallel transporting them along the $y$-direction. The total action is given by 

\begin{equation}\label{action-d}
S=2(S_{\text{\tiny NG}}+S_{\text{q}}\vert_{r=r_{\bar q}})
\,,
\end{equation}
where the boundary term describes a light quark (antiquark) attached to the string endpoint in the bulk. 

We first extremize the action with respect to the string profiles. Since we are interested in the case in which $x$ depends only on $r$, we choose the gauge $\xi^1=y$ and $\xi^2=r$. For the geometry \eqref{metric}, the Nambu-Goto action is then 

\begin{equation}\label{NG1}
S_{\text{\tiny NG}}=\g Y\int dr\,\frac{\ep^{\s r^2}}{r^2}\sqrt{f^{-1}+(\partial_r x^i)^2} 
\,.
\end{equation}
Here $Y=\int dy$. From this, it follows that $x^i=const$, which represents a straight string stretched along the $r$-axis, is a solution to the equations of motion. 

As in \cite{astb}, we consider a constant tachyon background $\text{T}_0$. For the worldsheets shown in Fig.\ref{confs}, whose boundaries are lines in the $y$-direction, the boundary action written in the gauge $\tau=y$ takes the form    

\begin{equation}\label{Sqy}
S_{\text q}=R{\text T}_0 Y\,\frac{\ep^{\frac{\s}{2}\rq^2}}{\rq}
\,.
\end{equation}
One immediately recognizes this as the action of a point particle of mass ${\text T}_0$ at rest.

Combining these results, the corresponding pseudopotential $\Vd=S/Y$ is 

\begin{equation}\label{Vd}
\Vd=
2\g\Bigl(
\int_0^{\rqb}\frac{dr}{r^2}\frac{\ep^{\s r^2}}{\sqrt{f}}
+
\n\frac{\ep^{\frac{\s}{2}\rqb^2}}{\rqb}
\Bigr)
\,,
\end{equation}
 where $\n=\frac{R{\text T}_0}{\g}$. The integral diverges at $r=0$ and therefore requires regularization. As in the case of the connected configuration \cite{az2}, we implement this by imposing a cutoff $\epsilon$ on the lower limit of integration

 \begin{equation}\label{regularization}
\int_{\epsilon}^{\rqb}\frac{dr}{r^2}\frac{\ep^{\s r^2}}{\sqrt{f}}
=
\frac{1}{\epsilon}-\frac{1}{\rqb}+\int_{\epsilon}^{\rqb}\frac{dr}{r^2}\Bigl(\frac{\ep^{\s r^2}}{\sqrt{f}}-1\Bigr)
\,.
\end{equation}
Subtracting the $\frac{1}{\epsilon}$ term and then taking $\epsilon\to 0$, we obtain the renormalized pseudopotential 

\begin{equation}\label{Vd2}
\Vd=
2\g\sqrt{\frac{\s}{\bar q}}\biggl(
\int_0^{1}\frac{dv}{v^2}
\biggl[\ep^{\bar q v^2}\biggl(1-\bar q^2v^4\frac{T^4}{T_c^4}\biggr)^{-\oh}-1-v^2
\biggr]\,
+
\n\ep^{\frac{\bar q}{2}}
\biggr)
+2c
\,,
\end{equation}
 where $\bar q=\s\rqb^2$. Importantly, the normalization constant $c$ is the same as in \eqref{El}. 

We still have to extremize the pseudopotential, or equivalently the action, with respect to $\rqb$. A simple calculation gives  

\begin{equation}\label{qb}
\ep^{\frac{\bar q}{2}}+\n(\bar q-1)\sqrt{1-\bar q^2\frac{T^4}{T_c^4}}
=0
\,.
\end{equation}
This equation determines $\bar q$ as a function of $T$. Its physical interpretation is as follows. It is the force balance equation at the light quark position. The two terms come from the gravitational force acting on the quark and the string tension at its position. Importantly, Eq.\eqref{qb} makes sense only if the light quarks are closer to the boundary than the soft wall and horizon. 

Thus, the pseudopotential associated with the disconnected configuration is given by the expression \eqref{Vd2} evaluated on the solution to Eq.\eqref{qb}. Clearly, it is independent of the length $\ell$. 

%________________________________________________________________________ 
\subsection{Details for $SU(3)$ gauge theory}

At this point, it makes sense to ask which configuration dominates in the sum \eqref{wilson2}. However, before answering this question, we need to specify the model parameters. In doing so, we use the same parameter set as in the case of the heavy quark-antiquark potential for $SU(3)$ gauge theory with two light quarks of equal mass \cite{astb}. First, the value of $\s$ is fixed from the slope of the Regge trajectory of $\rho(n)$ mesons in the soft wall model with the geometry \eqref{metric}, giving $\s=0.450\,\text{GeV}^2$ \cite{aq2}. Then, using the expression for $\sigma$, we obtain $\g=0.176$ by fitting the string tension to its value in \cite{bulava}.\footnote{Note that fitting the coefficient $\alpha$ in \eqref{small2} to the L\"usher value $\frac{\pi}{12}$ gives $\g=0.182$ that is quite satisfactory.} Finally, the parameter $\n$ is adjusted to reproduce the lattice result for the string breaking distance $\ell_c$ \cite{bulava}. With $\ell_c=1.22\,\text{fm}$, this gives $\n=3.057$.

\begin{figure}[htbp]
\centering
\includegraphics[width=6.9cm]{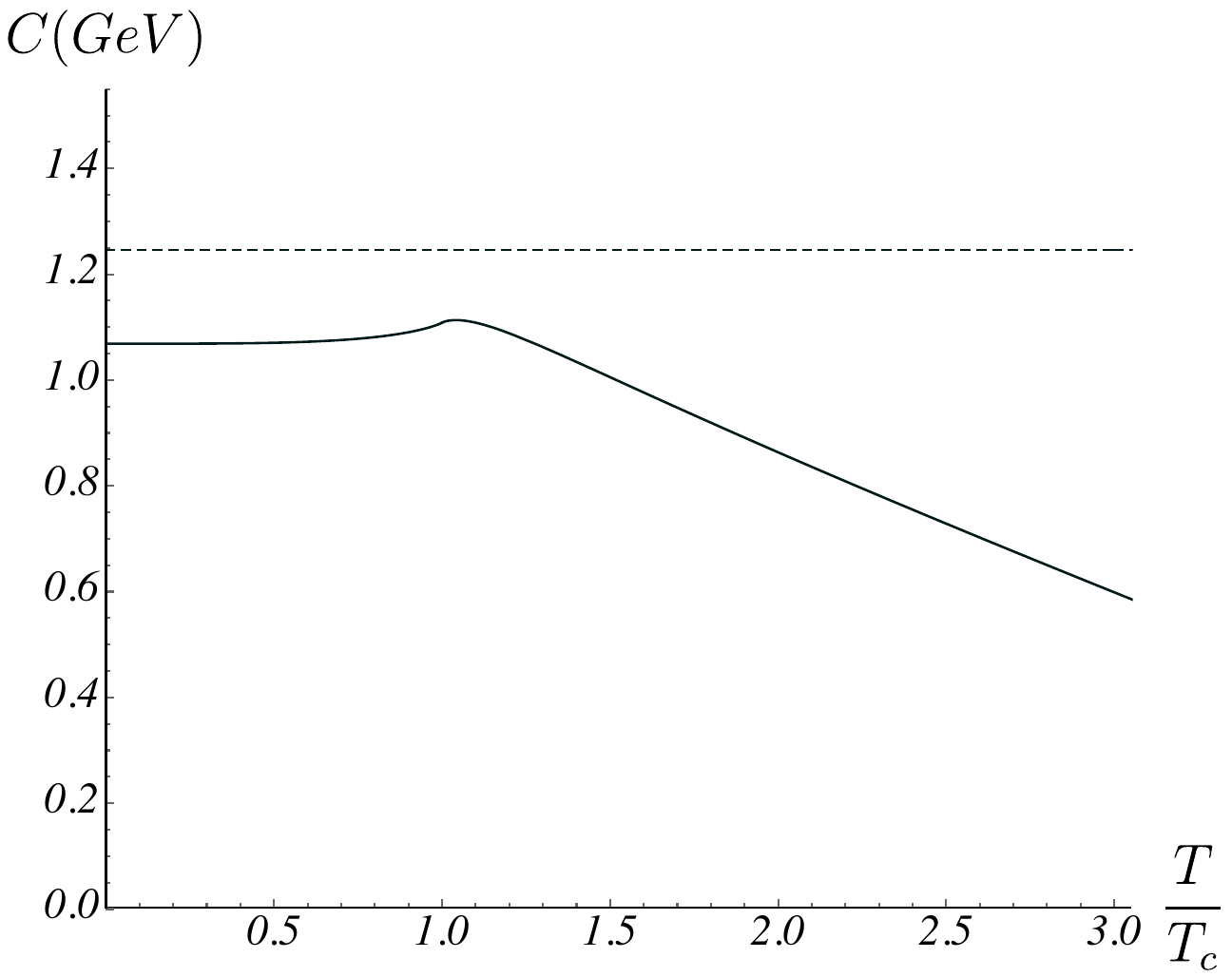}
\hspace{2.25cm}
\includegraphics[width=6.9cm]{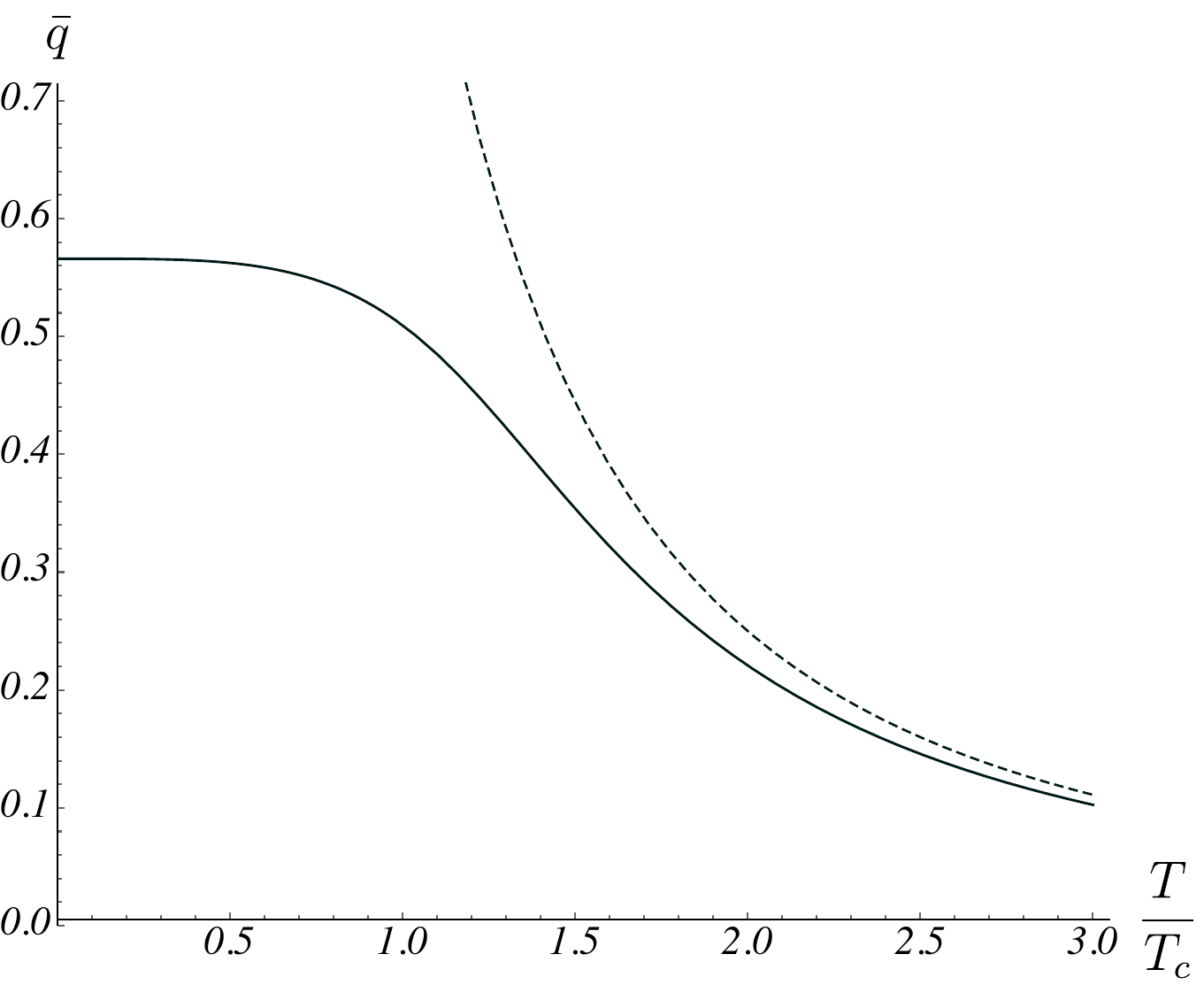}
\caption{{\small Left: The constant terms $C$ (solid) and $2c$ (dashed) as a function of temperature. We set $c=0.623\,\text{GeV}$, here and below. Right: The solution of Eq.\eqref{qb}. The dashed curve corresponds to the horizon in units $\s\rh^2$. The soft wall is at $1$ in these units.}}
\label{Cq}
\end{figure}

We are now in position to complete our discussion of the string configurations. Let us begin with the constant terms $C$ and $2c$. The results are shown on the left panel of Fig.\ref{Cq}. It is quite remarkable that the constant term $C$ also 
%________________________  fig - 2 __________________________________
%___________________________________________________________________
exhibits the specific behavior, like the spatial string tension. Namely, at low temperatures, it remains nearly constant; it then begins to slightly grow near the transition point, reaches a maximum (visible as a small bump at $T=1.04\,T_c$), and finally decreases with temperature.\footnote{Note that $C=2c-2\g\sqrt{\s}\frac{T}{T_c}+O\bigl(\frac{T_c^2}{T^2}\bigr)$ at high temperatures.} It is worth noting that the inequality $C<2c$ holds at all temperatures. In other words, the constant term in the large-$\ell$ expansion is smaller than that in the small-$\ell$ expansion. 

\begin{figure}[htbp]
\centering
\includegraphics[width=7.5cm]{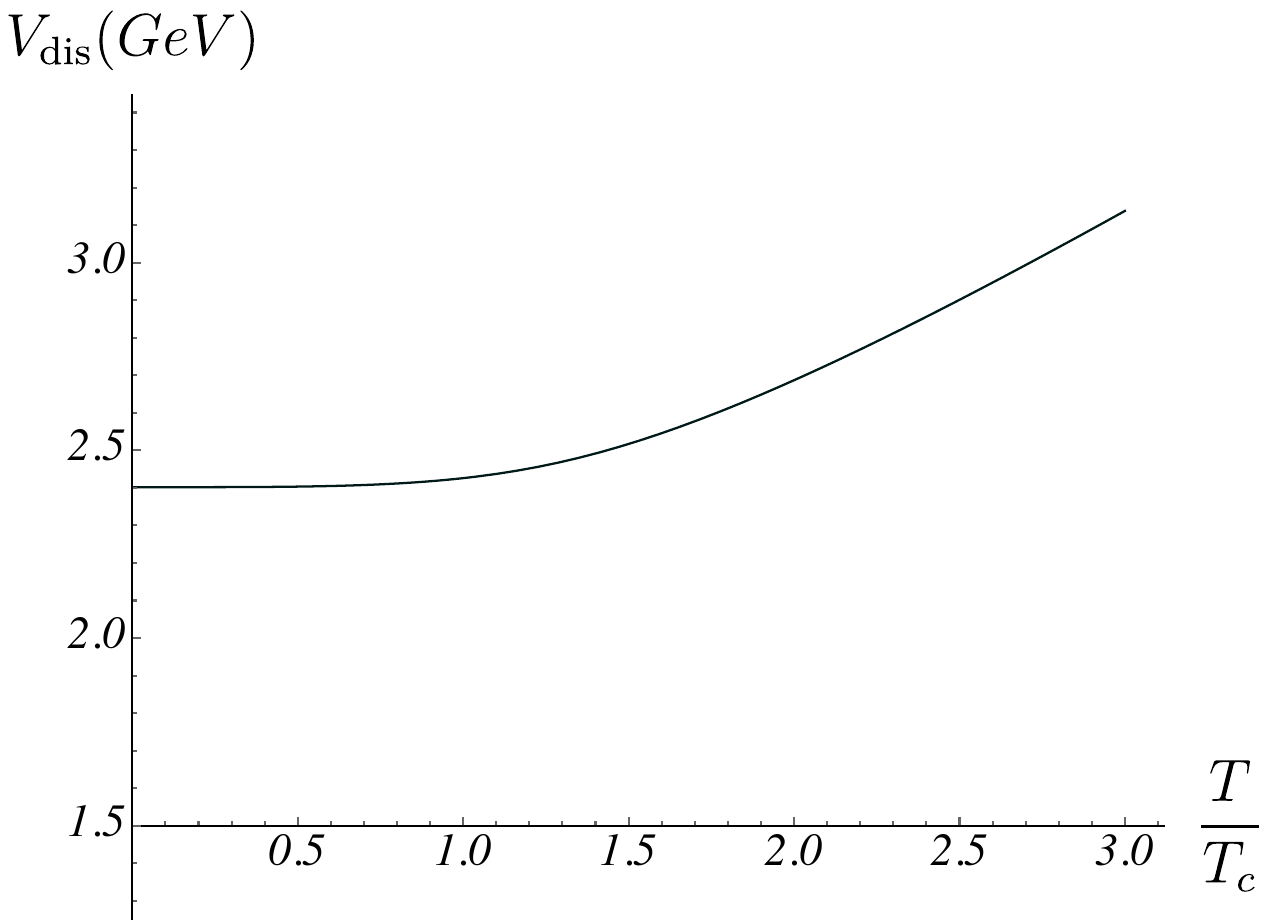}
\hspace{1.5cm}
\includegraphics[width=8.25cm]{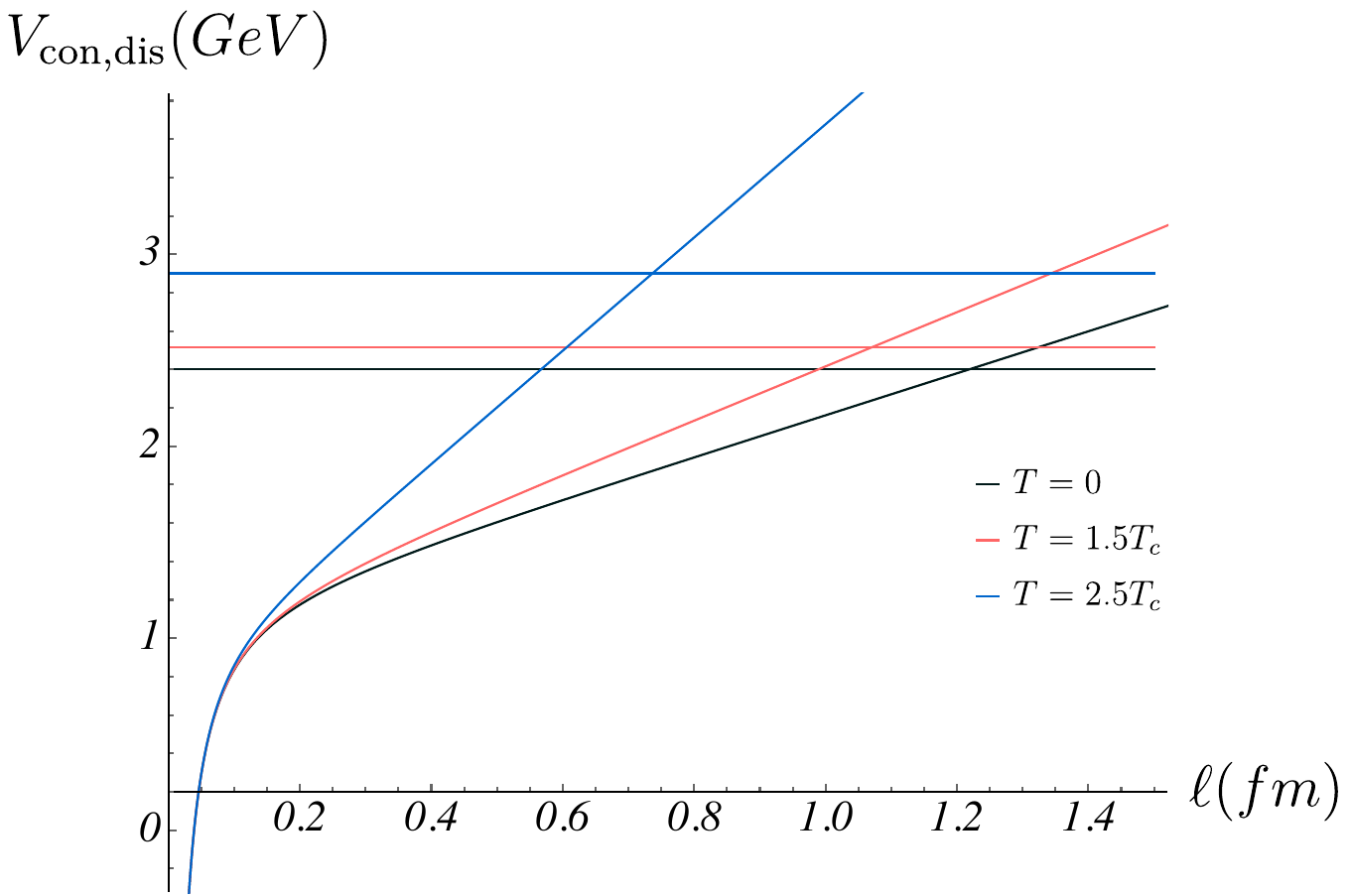}
\caption{{\small Left: $\Vd$ vs $\frac{T}{T_c}$. Right: $\Vc$ and $\Vd$ vs $\ell$ at various temperatures.}}
\label{VcVd}
\end{figure}

The equation \eqref{qb} can be solved numerically, with the result shown on the right panel of Fig.\ref{Cq}. The solution also changes its behavior across the transition point, but it does so quite smoothly. Clearly, the light quarks are outside the horizon. This remains true at higher temperatures, where $\bar q$ falls as $\sqrt{1-\n^{-2}}\,\frac{T^2_c}{T^2}$, while the horizon as $\frac{T_c^2}{T^2}$. With this solution, it is easy to find the temperature dependence of $\Vd$. For our parameter set, the result is presented on the left panel of Fig.\ref{VcVd}. We observe the specific behavior: the pseudopotential $\Vd$ remains nearly constant at low temperatures, then 
%________________________  fig - 3 __________________________________
%___________________________________________________________________
smoothly transitions near the critical point, and exhibits noticeable growth at higher temperatures. This growth becomes linear for $T\gtrsim 5T_c$, where $\Vd=\oh\g\sqrt{\s}\bigl(B(1-\n^{-2},-\tfrac{1}{4},\frac{1}{2})+\frac{4\n^2}{\sqrt{\n^2-1}}\bigr)\frac{T}{T_c}+2c+o(1)$. Here $B(z,a,b)$ denotes the incomplete beta function. 

We are now in a position to proceed with the pseudopotentials $\Vc$ and $\Vd$. On the right panel of Fig.\ref{VcVd}, we plot both as a function of the length $\ell$.  $\Vc$ exhibits the temperature-dependent behavior at large lengths and the temperature-independent behavior at small lengths, in agreement with the lattice results \cite{Vlattice}. Clearly, $\Vc$ dominates the pseudopotential $V$ at small lengths, whereas $\Vd$ dominates at large ones. Thus, the sum in Eq.\eqref{wilson2} involves the two distinct string configurations. This implies that the pseudopotential is given by $V=\min\{\Vc,\Vd\}$. 

\begin{figure}[htbp]
\centering
\includegraphics[width=8.15cm]{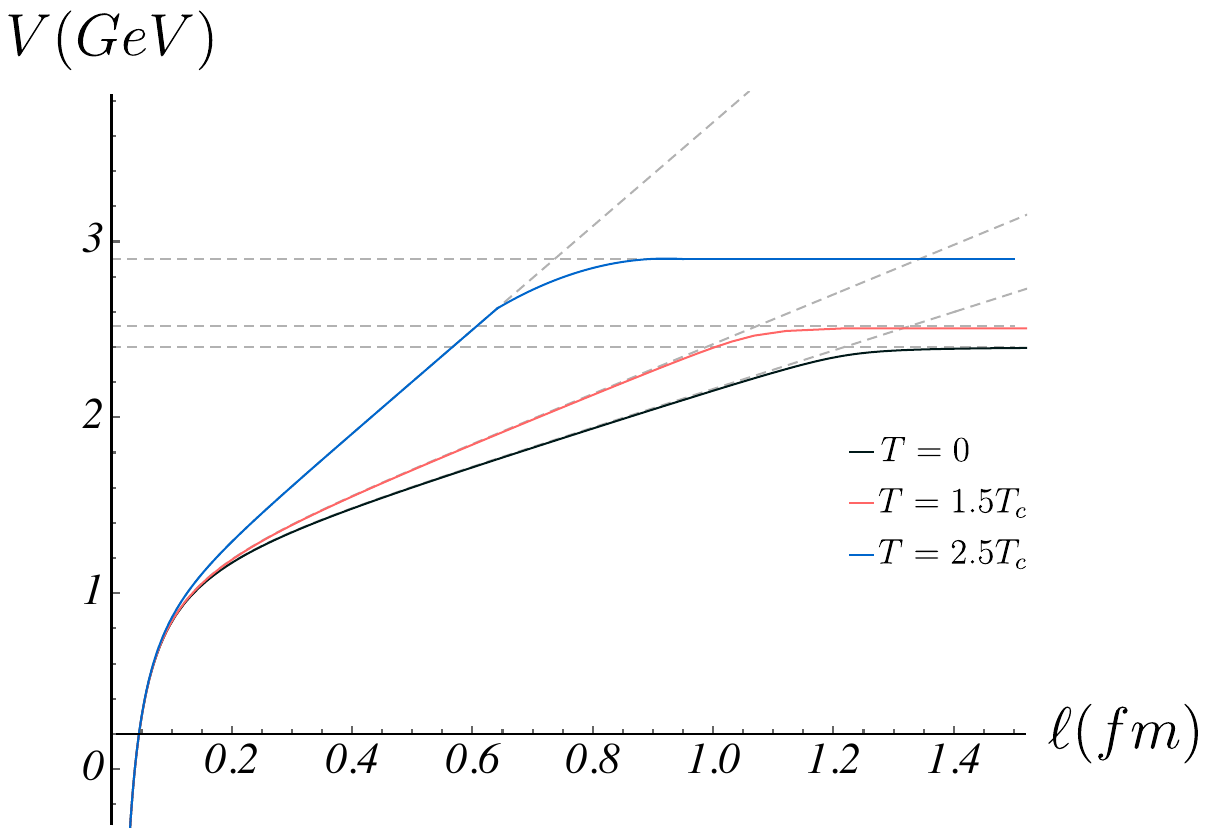}
\hspace{1.5cm}
\includegraphics[width=7.25cm]{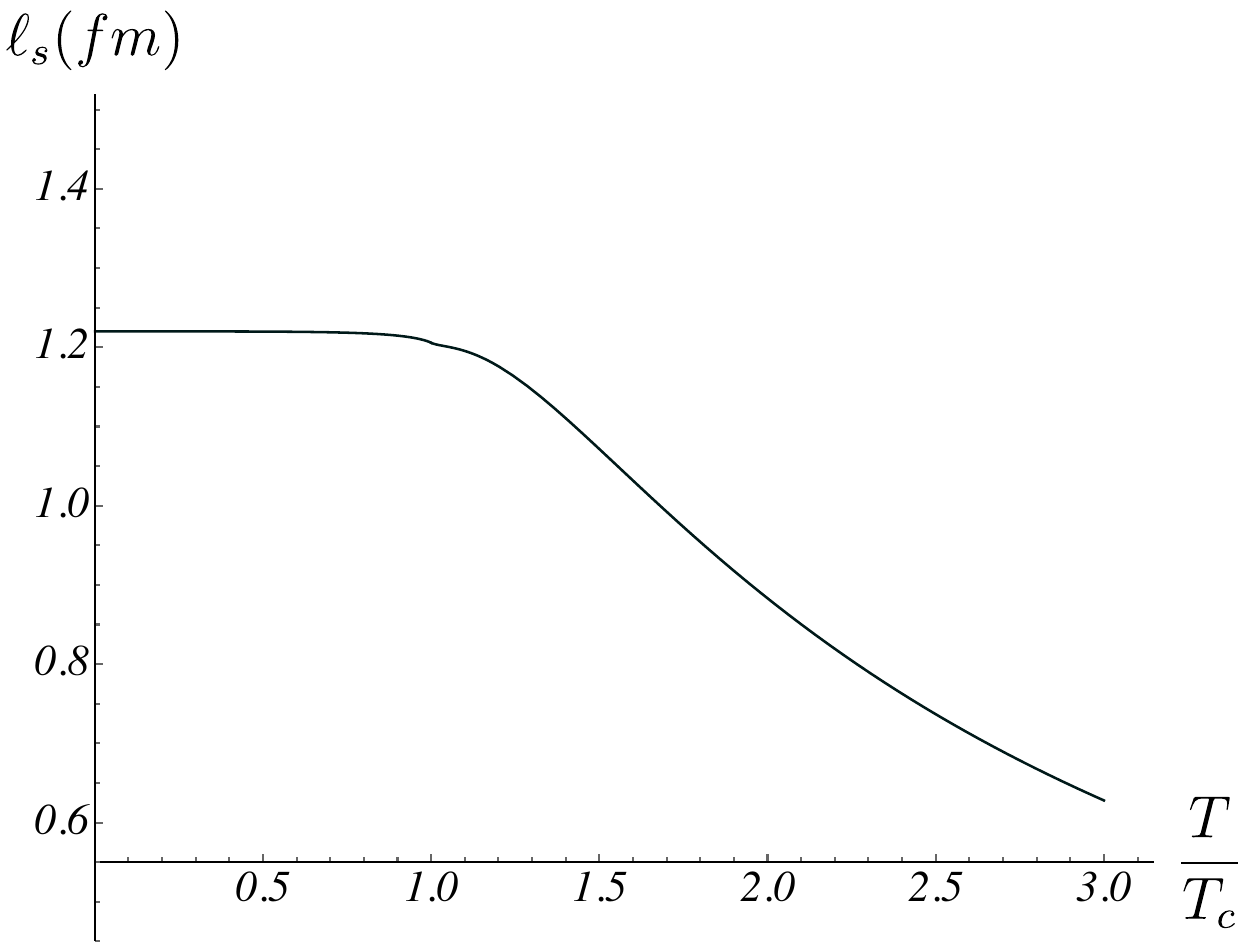}
\caption{{\small Left: Sketched here is the pseudopotential at various temperatures. $\Vc$ and $\Vd$ are shown in dashed lines. We set $\Theta=50\,\text{MeV}$ at $T=0$ and $T=1.5T_c$. Right: The spatial string breaking distance as a function of temperature.}}
\label{V}
\end{figure}

In practice, one particularly useful model, also used to determine the quark-antiquark potential \cite{bulava}, is that of \cite{drum}. It includes a mixing analysis based on a correlation matrix whose elements give rise to a model Hamiltonian. Adopting it for the problem in hand, we introduce  

\begin{equation}\label{H}
{\cal H}(\ell)=
\begin{pmatrix}
\Vc(\ell) & \Theta \\
\Theta & \Vd 
\end{pmatrix}
\,,
\end{equation}
where the diagonal elements represent the contributions from the string configurations, and the off-diagonal element describes the strength of mixing between them. The pseudopotential is then the smaller eigenvalue of ${\cal H}$. Explicitly,\footnote{The larger eigenvalue gives the pseudopotential $V_1$ (the subleading exponent in \eqref{Wlattice}) corresponding to the excited state.}

\begin{equation}\label{H2}
V=\oh\bigl(\Vc+\Vd\bigr)-\sqrt{\frac{1}{4}\bigl(\Vc-\Vd \bigr)^2+\Theta^2}
\,.
\end{equation}
Although we have computed the diagonal elements of ${\cal H}$, calculating the off-diagonal element within the effective string model remains challenging. Because of this, it is impossible to precisely visualize the form of the pseudopotential. However, we can still gain valuable insight from our experiences with the heavy quark-antiquark potential, particularly regarding the approximate magnitude of the $\Theta$ value.\footnote{For instance, one can assume that $\Theta$ is approximately constant, with a value around $50\,\text{MeV}$, as found on the lattice at zero temperature \cite{bulava}.} This leads to the overall picture sketched in Figure \ref{V} on the 
%________________________  fig - 4 __________________________________
%___________________________________________________________________
left. One important conclusion for what follows can be drawn from this result: the plots of $\Vc$ and $\Vd$ intersect at points near which $\Vc$ is approximately linear. 

To quantitively describe the flattening of the pseudopotential, we define the spatial string breaking distance $\ell_s$ by equating $\Vc$ and $\Vd$

\begin{equation}\label{ls}
	\Vc(\ell_s)=\Vd
	\,.
\end{equation}
At zero temperature, this definition coincides with that of the string breaking distance introduced in \cite{bulava} to characterize the heavy quark potential in the presence of light dynamical quarks. As a result, $\ell_s\vert_{T=0}=\ell_c$. Importantly, such defined $\ell_s$ is finite and scheme independent (i.e., independent of the normalization constant $c$). On the right panel in Fig.\ref{V}, we plot $\ell_s$ against temperature. The spatial string breaking distance weakly decreases at low temperatures but continues to decrease noticeably across the transition point. Thus, it also shows the specific behavior. By contrast, the string breaking distance $\ell_c$ increases weakly near zero temperature but rises more noticeably as temperature approaches the critical value \cite{astb}, becoming meaningless beyond the transition point.

The Eq.\eqref{ls} can be solved approximately if $\Vc$ is nearly linear. In this case, with the help of the asymptotic expansion \eqref{large-l6}, we find that 

\begin{equation}\label{ls1}
\ell_s\approx\frac{1}{\sigma_s}
\bigl(
\Vd-C
\bigr)
\,.
\end{equation}
In fact, this is a rather accurate approximation. In the temperature range $0\leq T\leq 3\,T_c$, the difference between the solution of Eq.\eqref{ls} and its approximation \eqref{ls1} appears only in the third digit after the decimal point. It is worth noting that the specific temperature behavior of the spatial string breaking distance follows directly from this formula. Indeed, all the terms on the right hand side exhibit the specific behavior across the transition point. Another conclusion is that $\ell_s$ scales as $\frac{1}{T}$ at very high temperatures. 

%________________________________________________________________
\section{Concluding comments}
\renewcommand{\theequation}{4.\arabic{equation}}
\setcounter{equation}{0}

In the context of the spatial Wilson loops, we considered string breaking to gain insight into the nonperturbative features of QCD with two light flavors. No phenomenological or lattice predictions yet exist in this context, but the effective string model at our disposal enables some predictions. In particular, we estimated the spatial string breaking distance for $SU(3)$ gauge theory in the temperature range $0\,\text{-}\,3T_c$.

The use of the effective string model at high temperatures requires a caveat. The temperature dependence of the spatial string tension agrees with the lattice data only for temperatures below $2.5\,\text{-}\,3\,T_c$, at least for pure gauge theories \cite{az2}. The similar holds for the Coulomb coefficient $\alpha$ defined in \eqref{small2} which begins to decrease at higher temperatures and eventually approaches the L\"uscher value $\frac{\pi}{24}$.\footnote{See, for example, the last reference in \cite{Vlattice}.}
 
We treated the off-diagonal element $\Theta$ of the model Hamiltonian as a free parameter. It would be of further interest to develop a string theory technique that enables its direct computation. Meanwhile, it would also be interesting to compute the model Hamiltonian ${\cal H}$ using lattice QCD or the $3$-dimensional effective field theory.

%____________________________________________________________
\begin{acknowledgments}
 
\end{acknowledgments}
We would like to thank Michele Caselle, Peter Petreczky and Peter Weisz for useful communications and for comments on the manuscript. We also thank Dima Levkov for hospitality at ITMP, where part of this work was completed. 
%__________________       R E F s     ______________________

%____________________________________________________________________
\end{document}